\documentclass[%
 reprint,
 amsmath,amssymb,
 aps,
]{revtex4-1}

\usepackage{graphicx}
\usepackage{dcolumn}
\usepackage{bm,url}

\begin{document}

\preprint{APS/123-QED}

\title{Hybrid particle-phase field model and renormalized surface tension in dilute suspensions of nanoparticles}

\author{Alexandra J. Hardy}
\author{Abdallah Daddi-Moussa-Ider}
\author{Elsen Tjhung}
\email{elsen.tjhung@open.ac.uk}
\affiliation{School of Mathematics and Statistics, The Open University, Walton Hall, Milton Keynes, MK7 6AA, United Kingdom}

\date{\today}

\begin{abstract}
We present a two-phase field model and a hybrid particle-phase field model to simulate dilute colloidal sedimentation and flotation near a liquid-gas interface (or fluid-fluid interface in general). 
Both models are coupled to the incompressible Stokes equation, which is solved numerically using a combination of sine and regular Fourier transforms to account for the no-slip boundary conditions at the boundaries. 
The continuum two-phase field model allows us to analytically solve the equilibrium interfacial profile using a perturbative approach, 
demonstrating excellent agreement with numerical simulations. 
Notably, we show that strong coupling to particle dynamics can significantly alter the liquid-gas interface, thereby modifying the liquid-gas interfacial tension.
In particular, we show that the renormalized surface tension is monotonically decreasing with increasing colloidal particle concentration and decreasing buoyant mass.
\end{abstract}
\maketitle


\section{Introduction}

Simulating the behaviour of colloidal particles near or at a fluid-fluid interface is a fascinating and complex area of research in soft matter~\cite{myers1999surfaces,binks2017colloids}.
The challenge arises from the need to accurately capture the multi-scale interactions and dynamics involved. 
This includes the colloidal particle-interface interactions, the hydrodynamics of the surrounding fluid, the influence of thermal fluctuations, and the presence of external forces such as gravity.
At the same time, there is still an ongoing debate whether colloidal particles near interfaces can also decrease the surface tension~\cite{dekker2023difference}, 
similar to surfactants~\cite{ji2020interfacial,zong2020modeling,van2006diffuse}.
Some studies report a decrease in surface tension with increasing colloidal particle concentration~\cite{okubo1995surface,dekker2023difference}, while some other studies indicate a non-monotonic behaviour~\cite{ranjbar2015experimental,dong2003surface,tanvir2012surface}. 
Furthermore, the effect of buoyant force on the surface tension remains understudied.

In previous simulations of colloidal suspensions~\cite{ladd2002lubrication}, the fluid flow is solved using the lattice Boltzmann algorithm, 
while the colloidal particles are modelled as solid nodes within the lattice. 
This approach accounts for the excluded volume interaction between the particles. 
In the context of spinodal phase separation, these particles tend to accumulate and jam at the interfaces, thereby arresting the phase separation~\cite{stratford2005colloidal,tiribocchi2019curvature}. 
However, this simulation method is computationally intensive~\cite{bonaccorso2020LB}, and the incompressibility condition in the lattice Boltzmann method is not strictly maintained~\cite{kruger2017LB}.

In the dilute limit, the excluded volume interaction between the particles can be neglected and the particles interact solely through hydrodynamic velocity, as demonstrated in previous studies~\cite{verberg2005modeling,balazs2000multi,ma2012modeling,tayeb2021evaporation}.
The purely diffusive case, without hydrodynamic flow, has also been considered in~\cite{yang2020phase-field,kim2020numerical}.
In the former scenario, the colloidal particles are treated as Brownian particles and the interaction with the fluid is assumed to be a long-range exponential decay.

In this paper, we simplify the interaction between colloidal particles and the fluid to be purely local, 
which we justify through a microscopic derivation from Flory-Huggins theory. 
We show that for large enough interaction strength (compared to the temperature scale), 
the colloidal particles will be confined inside one of the two fluid phases, \emph{e.g.} the liquid phase.
Experimentally, this scenario might correspond to the sedimentation of colloidal/nano-particles in a liquid solvent~\cite{midelet2017sedimentation}.
We also derive the continuum version of the model by taking an ensemble average of the stochastic particle dynamics. 
The continuum model offers the advantage of being solvable analytically using a perturbative approach, 
allowing us to determine the equilibrium liquid-gas interfacial profile and particle distribution, both of which show an excellent agreement with numerical simulations.

We demonstrate that with strong enough liquid-particle attraction, the presence of colloidal particles near the liquid-gas interface can significantly alter the interfacial profile, leading to a reduction in liquid-gas surface tension. 
This reduction in surface tension aligns with experimental observations reported in~\cite{okubo1995surface,dekker2023difference}.
Additionally, we predict that surface tension decreases with decreasing buoyant mass, though this phenomenon has not been extensively studied experimentally or theoretically.
This reduction in surface tension with increasing concentration is also observed in surfactant-laden interfaces, though the mechanism differs, as particles are adsorbed into the interface in the latter case, which gives rise to a different term in the free energy~\cite{xu2023improved}.

In our approach, the fluid flow is solved using a spectral method, strictly enforcing the incompressibility condition, unlike the lattice Boltzmann method in~\cite{verberg2005modeling,balazs2000multi,ma2012modeling}. 
We utilize a combination of regular Fourier transform along one axis and sine transform along the other axis to ensure no-slip boundary conditions at the walls  (which are parallel to one of the two axes).


\section{Models}

In this section, we will introduce two complementary models for describing particle sedimentation and flotation near a liquid-gas interface. 
The first model, the two-phase field model, is a purely continuum model that can be solved both analytically and numerically on a lattice. 
The second model is a hybrid approach that combines a continuum phase field with discrete point particles.

\begin{figure}[t]
    \centering
    \includegraphics[width=1.0\columnwidth]{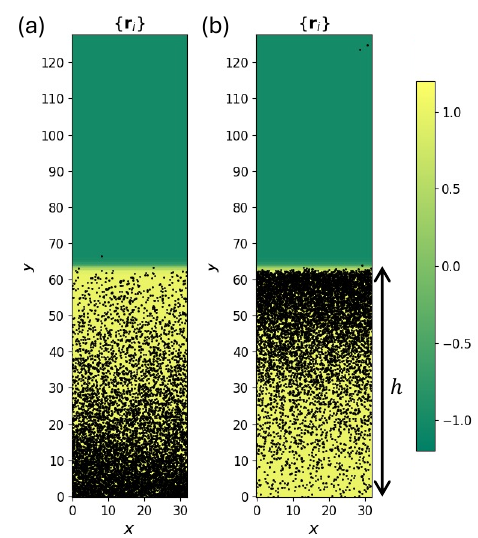}
    \caption{(a,b) show the steady state snapshots from the hybrid particle-phase field simulations for the parameter pairs $(C,G)=(3,0.05)$ and $(C,G)=(5,-0.05)$ respectively.  
    The color scale indicates the magnitude of $\phi(\mathbf{r},t)$, while the dots indicates the positions of the colloidal particles $\{\mathbf{r}_i\}$.
    $h$ is the height of the liquid-gas interface.
    (Parameters used: $B=500,M_\phi=0.02,\eta=10,L_x=32,L_y=128$, and $\Delta x=\Delta y=0.5$.)}
    \label{fig:2D}
\end{figure}

\subsection{Two-phase field model \label{sec:two-phase}}

In the two-phase field model~\cite{kim2012phase,mohamad2011lattice}, the hydrodynamic variables are: the fluid velocity $\mathbf{u}(\mathbf{r},t)$, the number density of the colloidal particles $\psi(\mathbf{r},t)$, and the combined number density of the colloidal particles plus the fluid `particles' $\phi(\mathbf{r},t)$ (see Appendix~\ref{app:lattice}).
Here, $\mathbf{u}(\mathbf{r},t)$, $\psi(\mathbf{r},t)$ and $\phi(\mathbf{r},t)$ are all functions of space $\mathbf{r}=(x,y,z)$ and time $t$, 
although in this paper we will focus mostly on the one-dimensional (1D) and two-dimensional (2D) case.
We assume a solid impenetrable wall at $y=0$, a flat liquid-gas interface at $y=h>0$, and the gravity is acting downwards in the negative $y$-direction, see Fig.~\ref{fig:2D}.

The combined number density $\phi(\mathbf{r},t)$ is rescaled and weighted by some factors such that $\phi$ is dimensionless and conserved, 
\emph{i.e.} $\int\phi(\mathbf{r},t)\,dV=$ constant for all $t$  (see Appendix~\ref{app:lattice} for more precise definition of $\phi$). 
By definition, $\phi>0$ represents the liquid phase and $\phi<0$ represents the gaseous phase. 
On the other hand, $\psi(\mathbf{r},t)$ is defined to be the unscaled number density of the colloidal particles.
$\psi$ has the dimension of one over volume.
$\psi$ is also conserved such that $\int\psi(\mathbf{r},t)\,dV=N_p$, where $N_p>0$ is the total number of colloidal particles in the system.
In most experimental situations, we assume the colloidal particles to be soluble in the liquid phase but not in the gaseous phase.
Thus, $\psi$ should be zero in the regions where $\phi<0$ and positive in the regions where $\phi>0$.
This `confinement' of $\psi$-field inside the positive regions of $\phi$-field is provided by a coupling constant $c>0$, which we shall introduce shortly below. 

Once we have defined all the hydrodynamic variables, we can then derive the equilibrium free energy $\mathcal{F}[\phi,\psi]$ as a functional of $\phi$ and $\psi$ (see Appendix~\ref{app:lattice}):
\begin{align}
\mathcal{F}[\phi,\psi] &= \int dV \bigg( \frac{\alpha}{2}\phi^2 + \frac{\beta}{4}\phi^4 + \frac{\kappa}{2}|\nabla\phi|^2 \nonumber\\
&+ \tilde{m}gy\psi + k_{\text{B}} T\psi\ln(a^3\psi) - c\phi\psi \bigg). \label{eq:free-energy}
\end{align}
$\alpha,\beta,\kappa$ and $c$ are thermodynamic constants, 
which are related to fluid-fluid and colloidal particle-fluid interaction microscopically.
For $\alpha<0$, the $\phi$-field will tend to phase separate into positive regions (corresponding to the liquid phase) and negative regions  (corresponding to the gaseous phase).
$\kappa$ is a phenomenological constant which can be approximated from the bare surface tension of the liquid-gas interface (\emph{i.e.} without any particle around).

The fourth term inside the integrand in Eq.~(\ref{eq:free-energy}) is the gravitational potential energy. 
$g$ is the acceleration of gravity and $\tilde{m}$ is the buoyant mass of the colloidal particles.
If $\tilde{m}>0$, the particles tend to sink, and if $\tilde{m}<0$, the particles tend to float.
The fifth term is the entropic term which originates from Brownian motion of the colloidal particles.
$k_{\text{B}}$ is the Boltzmann constant and $T$ is the temperature. 
$a$ is the typical size of the colloidal particles and we have introduced a factor of $a^3$ inside the logarithm since $\psi$ has the dimension of one over volume.
Finally the sixth term is the coupling term.
If $c>0$, the particles will be soluble in the liquid phase, and if $c<0$, the particles will be soluble in the gaseous phase (which is unphysical).
The coupled term $\propto c$ competes with the entropic term $\propto k_{\text{B}}T$ and we require $c\gg k_{\text{B}}T$ to fully confine the particles inside the liquid phase.
For example in Fig.~\ref{fig:2D}(b) for $C=c/k_{\text{B}}T=5$, two particles managed to escape into the gas phase, however compared to the total number of particles, 
this is statistically negligible.

The dynamics of $\phi(\mathbf{r},t)$ and $\psi(\mathbf{r},t)$ are then given by the advection-diffusion equations:
\begin{align}
\frac{\partial\phi}{\partial t} + \left(\mathbf{u}\cdot\nabla\right)\phi &= M_\phi\nabla^2\frac{\delta\mathcal{F}}{\delta\phi}, \label{eq:phidot}\\
\frac{\partial\psi}{\partial t} + \left(\mathbf{u}\cdot\nabla\right)\psi &= \frac{1}{\lambda}\nabla\cdot\left(\psi\nabla\frac{\delta\mathcal{F}}{\delta\psi}\right), \label{eq:psidot}
\end{align} 
where $M_\phi>0$ is a mobility constant and $\lambda>0$ is the friction coefficient between the colloidal particles and the liquid.
For spherical particles, $\lambda$ can be written as $\lambda=6\pi\eta R$, where $\eta$ is the viscosity of the liquid and $R$ is the radius of the particle.
The terms proportional to $\mathbf{u}$ in Eqs.~(\ref{eq:phidot}-\ref{eq:psidot}) represent advection of fluid and colloidal particles by the fluid velocity $\mathbf{u}$.
The terms of order $\sim\nabla^2$ in Eqs.~(\ref{eq:phidot}-\ref{eq:psidot}) represent diffusion of fluid and colloidal particles from high to low chemical potential.
$\delta\mathcal{F}/\delta\phi$ and $\delta\mathcal{F}/\delta \psi$ are the chemical potentials for $\phi$ and $\psi$ respectively.

Assuming small Reynolds number, the fluid velocity inside the liquid phase $\mathbf{u}(\mathbf{r},t)$ satisfies the incompressible Stokes equation:
\begin{equation}
0=-\nabla p + \eta\nabla^2\mathbf{u} + \mathbf{f}[\phi,\psi], \quad\text{and}\quad \nabla\cdot\mathbf{u}=0,
\label{eq:Stokes}
\end{equation}
where $\eta$ is the viscosity of the liquid and $p(\mathbf{r},t)$ is the pressure.  
$\mathbf{f}$ is the force density, which depends on $\phi$ and $\psi$, and can be written as a gradient of the elastic stress tensor: 
$\mathbf{f}=\nabla\cdot\underline{\underline{\boldsymbol{\sigma}}}$.
The expression for $\mathbf{f}$ (or equivalently $\underline{\underline{\boldsymbol{\sigma}}}$) can be derived from the free energy functional (\ref{eq:free-energy}) (see Appendix~\ref{app:stress}):
\begin{equation}
\mathbf{f}[\phi,\psi]=\nabla\cdot\underline{\underline{\boldsymbol{\sigma}}}[\phi,\psi] = -\phi\nabla\frac{\delta\mathcal{F}}{\delta\phi} - \psi\nabla\frac{\delta\mathcal{F}}{\delta\psi}. \label{eq:force-density}
\end{equation}
The fluid flow in the gaseous phase (outside the liquid phase) is turbulent and characteristic of a high Reynolds number flow. 
However, this turbulent flow is well separated from the smooth Stokesian flow within the liquid by a boundary layer near the liquid-gas interface. 
Therefore, it is standard practice to solve the Stokes equation (\ref{eq:Stokes}) for the entire region and disregard the solution for $\mathbf{u}(\mathbf{r},t)$ in the gaseous phase.
Note that gravity is acting on the colloidal particles through the buoyant mass $\tilde{m}$, and is present in the Stokes equation through $\mathbf{f}[\phi,\psi]$.

In the absence of colloidal particles ($\psi=0$) or when the dynamics are fully decoupled ($c=0$), the free energy in Eq.~(\ref{eq:free-energy}) reduces to the standard Ginzburg-Landau free energy for liquid-gas phase separation~\cite{cates2018review}.
The equilibrium state $t\rightarrow\infty$ is then given by the minimum of the free energy $\delta\mathcal{F}/\delta\phi=0$, which gives the interfacial profile for the fluid:
\begin{equation}
\phi(y) = \sqrt{\frac{-\alpha}{\beta}} \tanh\left(\frac{h-y}{\xi}\right), \,\,\text{where}\,\, \xi=\sqrt{\frac{-2\kappa}{\alpha}}. \label{eq:phi-classical}
\end{equation}
Thus in phase field modelling, the liquid-gas interface at $y=h$ is not sharp, but is instead diffuse with some interfacial width $\xi$
(assumed to be small compared to the system size).
In this limit, the liquid-gas surface tension can also be derived: $\gamma_0=\sqrt{-8\kappa\alpha^3/9\beta^2}$~\cite{cates2018review}.
We call $\gamma_0$ the bare surface tension. 
In Section~{\ref{sec:surface-tension}},
we will discuss how the surface tension is modified by the presence of colloidal particles near the interface.


\subsection{Hybrid particle-phase field model \label{sec:hybrid}}

Instead of a smooth density field $\psi(\mathbf{r},t)$, we may also represent the colloidal particles as point particles, whose positions are located at 
$\{\mathbf{r}_i(t); \text{ where } i=1,2,\dots,N_p\}$.
The dynamics of $\mathbf{r}_i(t)$ follows the overdamped Langevin equation:
\begin{equation}
\frac{d\mathbf{r}_i}{dt} = \mathbf{u}(\mathbf{r}_i,t) - \frac{\tilde{m}g}{\lambda} \hat{\mathbf{y}} + \frac{c}{\lambda}\nabla\phi(\mathbf{r}_i,t) + \sqrt{\frac{2k_{\text{B}} T}{\lambda}} \boldsymbol{\zeta}_i(t), \label{eq:rdot}
\end{equation}
where $\lambda>0$ is the friction coefficient and $\hat{\mathbf{y}}$ is a unit vector in the positive $y$-direction.
Note that the coupled term $\propto c$ appears as a force, which prevents the particles from escaping the interface.
$\boldsymbol{\zeta}_i(t)$ in (\ref{eq:rdot}) is a Gaussian white noise with zero mean and delta-correlation:
\begin{equation}
\left<\zeta_{i\alpha}(t)\zeta_{j\beta}(t')\right> = \delta_{ij}\delta_{\alpha\beta}\delta(t-t'), \label{eq:zeta}
\end{equation}
where the subscripts $\alpha$ and $\beta$ indicate the Cartesian coordinates $x,y,$ or $z$. 
The last term in Eq.~(\ref{eq:rdot}) is a random force acting on the particles, with a magnitude proportional to square root of temperature, as required by fluctuation-dissipation theorem~\cite{kubo1966fluctuation}.
In Eq.~(\ref{eq:rdot}), the random force is assumed to be additive, \emph{i.e.} the variance does not depend on position explicitly.
First principle derivation of the random force has suggested that the random force may become multiplicative for large enough shear flow~\cite{zaccone2023langevin}.

Eq.~(\ref{eq:rdot}) has to be solved together with the $\phi$-dynamics in Eq.~(\ref{eq:phidot}) and the Stokes equation in Eq.~(\ref{eq:Stokes}).
However, $\psi(\mathbf{r},t)$ appears explicitly in both (\ref{eq:phidot}) and (\ref{eq:Stokes}) and to solve these equations,
we need to coarse-grain all the particles' positions $\{\mathbf{r}_i(t)\}$ into the density field $\psi(\mathbf{r},t)$ via:
\begin{equation}
\psi(\mathbf{r},t) = \sum_{i=1}^{N_p} \delta(\mathbf{r} - \mathbf{r}_i(t)). \label{eq:coarse-grained-psi}
\end{equation}
In numerical simulations, say in 2D, the space $\mathbf{r}\in\mathbb{R}^2$ is discretized into a lattice of grid size $\Delta x\times \Delta y$.
Thus the summation over Dirac delta functions in (\ref{eq:coarse-grained-psi}) is practically just a histogram over all particles' positions into a 2D lattice grid.

Taking the ensemble average of Eq.~(\ref{eq:rdot}), the probability distribution for the particles' positions $P(\mathbf{r},t)$ is then given by the Fokker-Planck equation~\cite{vankampen2007stochastic}:
\begin{equation}
\frac{\partial P}{\partial t} + \nabla\cdot\left(P\mathbf{u} - \frac{\tilde{m}g}{\lambda}P\hat{\mathbf{y}} + \frac{c}{\lambda}P\nabla\phi - \frac{k_{\text{B}}T}{\lambda}\nabla P \right) = 0, \label{eq:Pdot}
\end{equation}
with normalization condition $\int P(\mathbf{r},t)\,dV=1$.
One can may then observe that Eq.~(\ref{eq:Pdot}) is identical to Eq.~(\ref{eq:psidot}) with the free energy functional given in Eq.~(\ref{eq:free-energy}) by replacing $P(\mathbf{r},t)=\psi(\mathbf{r},t)/N_p$.
Thus both particle-based description and purely continuum description of the system are statistically equivalent.

From (\ref{eq:rdot}) or (\ref{eq:psidot}), we can also estimate the time it takes for the system to reach sedimentation equilibrium from an initially uniformly distributed $\psi(y)$ in the region $y\in[0,h]$.
This is roughly given by the height of the interface $h$ divided by the sedimentation velocity $\tilde{m}g/\lambda$~\cite{midelet2017sedimentation}:
\begin{equation}
t_\text{eq} = \frac{h\lambda}{\tilde{m}g}. \label{eq:t-eq} 
\end{equation}
In a $1$ cm high liquid column, the time it takes for typical gold nanoparticles to reach sedimentation equilibrium varies from hours to weeks~\cite{midelet2017sedimentation}.

\section{Results \label{sec:analytic}}

From now on, we will take $\xi=\sqrt{-2\kappa\alpha/\beta}$ as the unit of length, $\tau=\lambda\xi^2/(k_{\text{B}}T)$ as the unit of time, and $k_{\text{B}}T$ as the unit of energy.
In other words, we fix $\xi=1$, $\lambda=1$ and $k_{\text{B}}T=1$.
In these units, the free energy (\ref{eq:free-energy}) can be expressed as (see Appendix~\ref{app:dimensionless}):
\begin{align}
\mathcal{F}[\phi,\psi] &= \int dV \bigg( -\frac{B}{2}\phi^2 + \frac{B}{4}\phi^4 + \frac{B}{4}|\nabla\phi|^2 \nonumber\\
&+ Gy\psi + \psi\ln\psi - C\psi\phi \bigg), \label{eq:F-dimensionless}
\end{align}
while the equations of motion remain the same.
Here $C$ is the coupling constant in units of $k_{\text{B}}T$ and $G$ is the buoyant mass in units of $k_{\text{B}}T/(g\xi)$ (see Appendix~\ref{app:dimensionless}).
For $G>0$, the particles will tend to sink, and for $G<0$, the particles will tend to float.
In the absence of colloidal particles $\psi(y)=0$, minimisation of (\ref{eq:F-dimensionless}) will give the classical result for liquid-gas interface: $\phi_0(y)=\tanh(h-y)$ with the bare surface tension: $\gamma_0=2B/3$ (in units of $k_{\text{B}}T/\xi^2$).
In these units, the Langevin equation for the hybrid particle-phase field model Eq.~(\ref{eq:rdot}) becomes:
\begin{align}
\frac{d\mathbf{r}_i}{dt} = \mathbf{u}(\mathbf{r}_i,t) - G \hat{\mathbf{y}} + C\nabla\phi(\mathbf{r}_i,t) + \sqrt{2} \boldsymbol{\zeta}_i(t),
\end{align}
where $\boldsymbol{\zeta}_i(t)$ is the dimensionless Gaussian white noise with zero mean and delta correlation, \emph{cf.} Eq.~(\ref{eq:zeta}).

\subsection{Perturbative equilibrium solution \label{sec:analytic}}

We will now derive analytically the equilibrium interfacial profile $\phi(y)$ and particle density $\psi(y)$ using a perturbative approach for a semi-infinite system $y\in[0,\infty)$.
The result is shown to match with the numerical simulations as long as the system size is large enough compared to the correlation length $\xi$.
Note that in equilibrium, the fluid velocity will be zero $\mathbf{u}=0$.
To find the equilibrium densities $\phi(y)$ and $\psi(y)$, we set the left hand sides of Eqs.~(\ref{eq:phidot}-\ref{eq:psidot}) to zero.
Using the free energy expression in (\ref{eq:F-dimensionless}) and imposing the boundary conditions $\phi(y\rightarrow\infty)=-1$ (gaseous phase at infinity) and $\psi(y\rightarrow\infty)=0$ (no particle at infinity), Eqs.~(\ref{eq:phidot}-\ref{eq:psidot}) become:
\begin{align}
	-\phi(y) + \phi(y)^3 - \tfrac{1}{2} \phi''(y) - \epsilon\psi(y) &= 0, \label{eq:phidot-eq} \\[3pt]
	\psi'(y) + G \psi(y) - C \psi(y) \phi'(y) &= 0. \label{eq:psidot-eq}
\end{align}
Here, $\epsilon = C/B \ll 1$ represents a small parameter (in the simulations $\epsilon\sim0.01$).
To solve Eq.~\eqref{eq:phidot-eq} for $\phi(y)$, we will employ a perturbative approach.
Let us set:
\begin{equation} 
\phi(y) = \phi_0(y) + \epsilon \phi_1(y) + \mathcal{O} \left(\epsilon^2 \right). \label{eq:sol_phi}
\end{equation}
Substituting this expression into Eq.~\eqref{eq:phidot-eq}, we obtain:
\begin{align}
	-\phi_0(y) + \phi_0(y)^3 - \tfrac{1}{2} \phi_0''(y) &= 0 \quad\text{and} \label{eq:phi0} \\
	-\phi_1(y) + 3 \phi_0(y)^2 \phi_1(y) - \tfrac{1}{2} \phi_1''(y) &= \psi(y), \label{eq:phi1}
\end{align}
at order $\epsilon^0$ and $\epsilon^1$ respectively.
The solution of Eq.~\eqref{eq:phi0} for \( \phi_0(y) \) is represented by the classical hyperbolic tangent solution, \emph{cf.} Eq.~\eqref{eq:phi-classical}, 
$\phi_0(y) = \tanh(h-y)$.
Directly solving Equation~\eqref{eq:psidot-eq} results in,
\begin{equation}
	\psi(y) = c_1 \exp \left[ -Gy + C \phi(y) \right],
	\label{eq:sol_psi}
\end{equation}
where $c_1$ is a constant, determinable from the normalization condition $\int_0^\infty\psi(y)\,dy=N_p$.

To facilitate analytical progress in solving Eq.~\eqref{eq:phi1} for $\phi_1(y)$, we will resort to the following approximation
\begin{equation}
	\phi_0(y) \approx H(h-y) - H(y-h),
\end{equation}
where $H(y)$ is the Heaviside step function so that $\phi_0(y)^2 \approx 1$ in Eq.~\eqref{eq:phi1}.
We can then solve \eqref{eq:phi1} to find
\begin{align}
	\phi_1^\pm(y) &= c_2^\pm \exp ( -2y ) + c_3^\pm \exp ( 2y ) \notag \\
	&\quad+ \tfrac{1}{2} \, \exp (-2y) \int \exp(2y) \psi^\pm(y) \, dy \notag \\
	&\quad- \tfrac{1}{2} \, \exp (2y) \int \exp(-2y) \psi^\pm(y) \, dy \, ,
	\label{eq:sol_phi1} 
\end{align}
where the superscripts $-$ and~$+$ pertain to the regions below and above the interface, respectively.

By approximating \(\psi^\pm(y) \approx c_1 \exp \left( -Gy \mp C \right)\), we obtain,
\begin{align}
	\phi_1^{\pm}(y) &= c_2^\pm \exp ( -2y ) + c_3^\pm \exp ( 2y ) \notag \\
	&\quad+ \frac{2 c_1}{4-G^2} \,  \exp \left( -Gy \mp C \right). \label{eq:sol_phi1}
\end{align}
It follows from the regularity condition at infinity that $c_3^+ = 0$.
For the boundary conditions, we demand
\begin{align}
		\left. \partial_y \phi_1(y) \right|_{y = 0} &= 0, \label{eq:BC-1} \\
		\left. \phi_1^+(y) - \phi_1^-(y) \right|_{y=h} &= 0, \label{eq:BC-2} \\
		\left. \partial_y \phi_1^+(y) - \partial_y \phi_1^-(y) \right|_{y=h} &= 0. \label{eq:BC-3}
\end{align}
The boundary condition in (\ref{eq:BC-1}) comes from the `neutral' wetting condition (see Appendix~\ref{app:numerics} and~\cite{kruger2017LB}),
while (\ref{eq:BC-2}-\ref{eq:BC-3}) come from the continuity condition at $y=h$.
This leads to the determination of $c_2^\pm$ and $c_3^-$ as
\begin{align}
	\frac{c_2^+}{c_1} &= \left( \frac{ \exp \left[ (2-G)h\right] }{2-G} - \frac{ \exp \left[ -(2+G)h\right] }{2+G} \right) \sinh (C) \notag \\
				   &\quad-  \frac{G \exp(C)}{4-G^2} \, , \\
	\frac{c_2^-}{c_1} &= -\frac{\sinh (C)}{2+G} \, \exp \left[- (2+G)h\right] - \frac{G \exp(C)}{4-G^2} \, , \\
	\frac{c_3^-}{c_1} &= -\frac{\sinh (C)}{2+G} \, \exp \left[- (2+G)h\right] \, .
\end{align}
The perturbative solution in Eq.~\eqref{eq:sol_phi1} should also work for the case of flotation ($G<0$). 
In this case, $\phi_1(y)$ diverges as $y\rightarrow\infty$.
However in numerical simulations, the system is finite $y\in[0,L]$, and as long as $L\gg G^{-1}$, the number of particles escaping from the liquid-gas interface is statistically negligible. 

\begin{figure}
    \centering
    \includegraphics[width=1\linewidth]{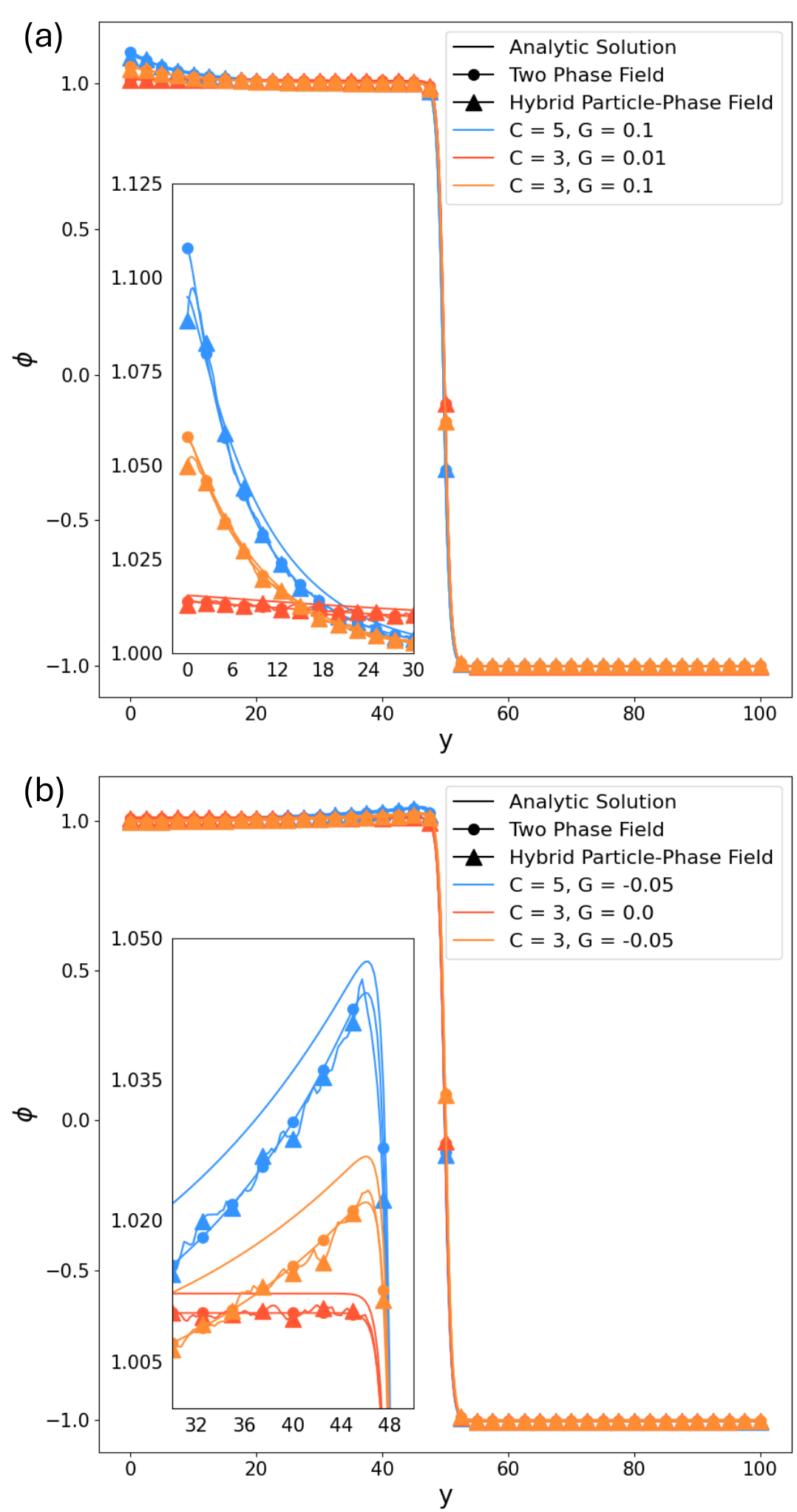}
    \caption{(a) shows the interfacial profile $\phi(y)$ for different values of $C$ and $G>0$.
    Triangles are the results from hybrid particle-phase field simulation, circles are the results from two-phase field simulations, and solid lines are the prediction from perturbation theory.
    (b) shows similar plots for different values of $G\le0$. 
    (Parameters used: $B=500,M_\phi=0.002,L_x=1,L_y=100,\Delta x=1$ and $\Delta y=0.25$.)}
    \label{fig:phi}
\end{figure}

\begin{figure*}
    \centering
    \includegraphics[width=1\textwidth]{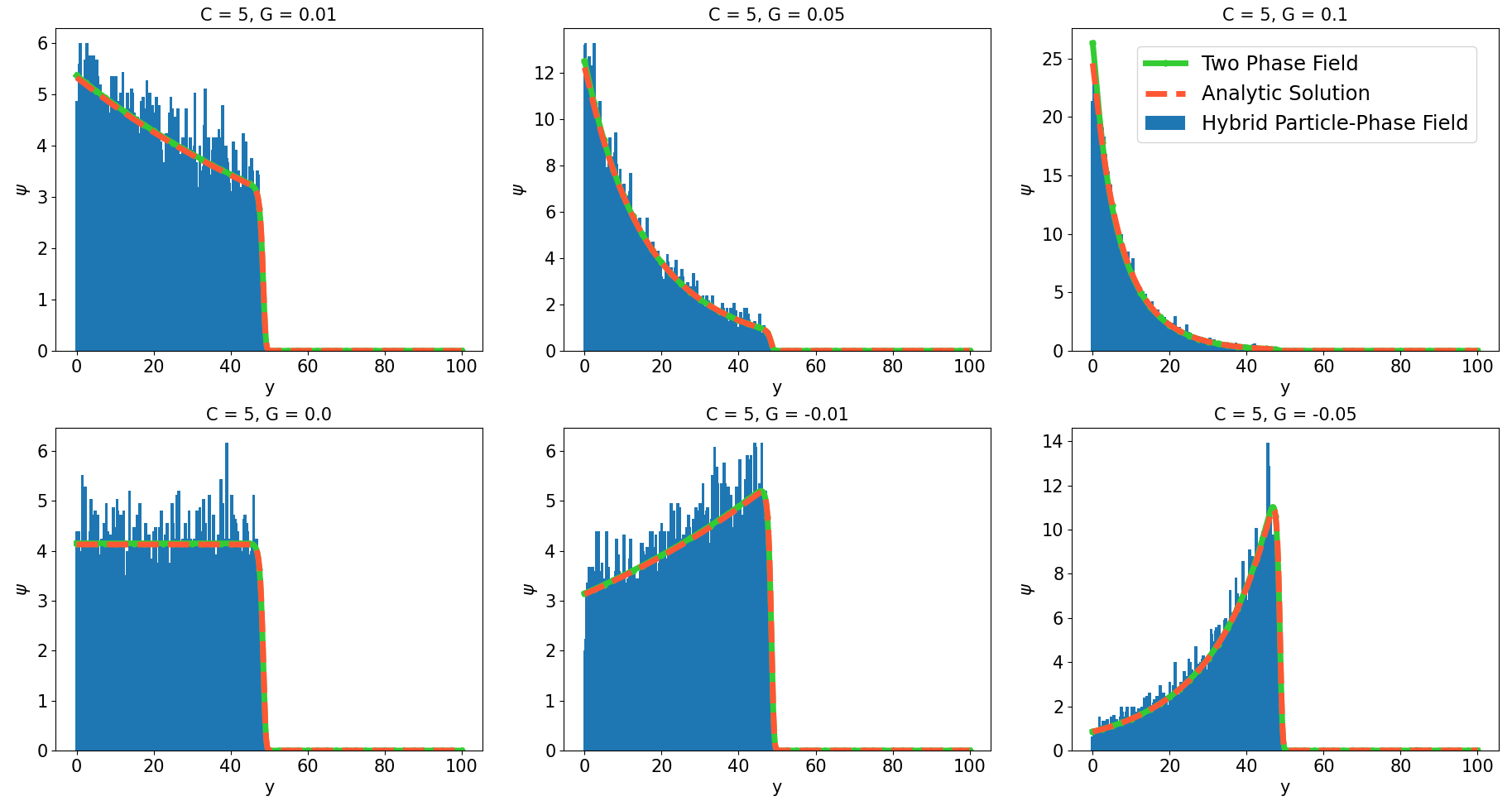}
    \caption{Blue show the histograms of particles' $y$-positions from hybrid particle-phase field simulation for different values of buoyant mass $G$, averaged over $50$ simulation runs.
    Green lines show particle density $\psi(y)$ from two-phase field simulation. 
    Orange lines show the prediction of $\psi(y)$ from perturbation theory. 
    (Parameters used: $B=500,M_\phi=0.002,L_x=1,L_y=100,\Delta x=1$ and $\Delta y=0.25$.)}
    \label{fig:psi}
\end{figure*}

Finally, the perturbative solutions are presented as solid lines in Fig.~\ref{fig:phi} and orange curves in Fig.~\ref{fig:psi}, 
showing an excellent agreement with numerical simulations.

\subsection{Renormalized surface tension \label{sec:surface-tension}}

We can now investigate how the presence of colloidal particles in the liquid phase modifies the liquid-gas surface tension.
From Appendix~\ref{app:stress}, we know that $\psi$ only appears explicitly as an isotropic pressure term in the Stokes equation.
Consequently, $\psi$ does not directly influence the surface tension of the liquid-gas interface. 
However as we can see from the previous section, $\psi(y)$ can alter the interfacial profile $\phi(y)$, which in turn affects the surface tension.

\begin{figure}
    \centering
    \includegraphics[width=1\linewidth]{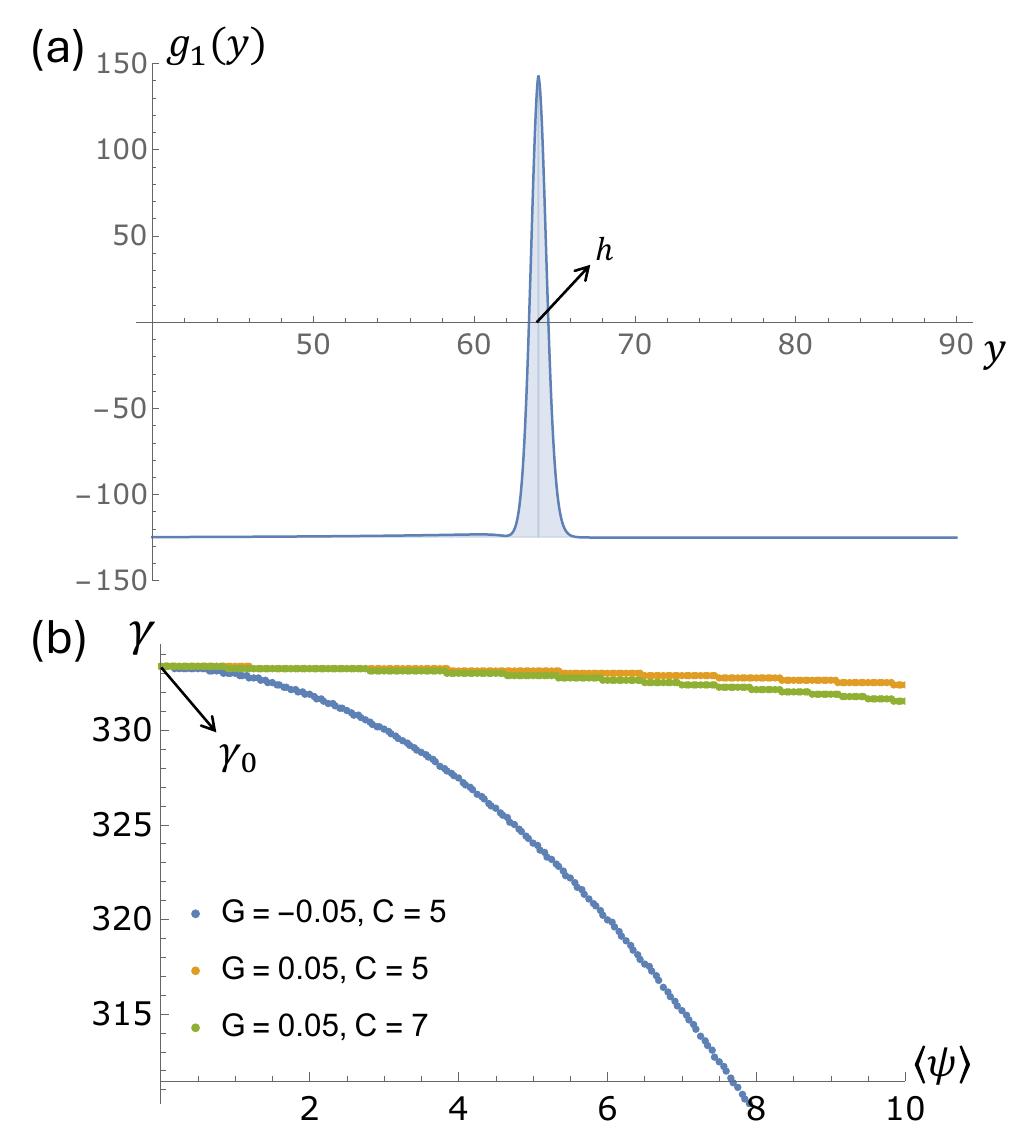}
    \caption{(a) shows the equilibrium free energy density $g_1(y)$ as a function of vertical distance $y$ for $\left<\psi\right>=4$ and $G=-0.05$.
    The surface tension is defined to be the excess free energy, which is the shaded blue area in the plot.
    (b) shows the surface tension as a function of average concentration of particles $\left<\psi\right>$ for $(C,G)=(5,-0.05)$ (blue), $(5,0.05)$ (orange) and $(7,0.05)$ (green).
    (Parameters used: $B=500$, and $h=64$.)}
    \label{fig:surface-tension}
\end{figure}

To see this, we substitute the equilibrium solution for $\psi(y)$ in Eq.~(\ref{eq:sol_psi}) into the free energy functional in Eq.~(\ref{eq:F-dimensionless}) to get the equilibrium free energy:
\begin{align}
\mathcal{F}_{\text{eq}} = \int dV \left( -\frac{B}{2}\phi^2 + \frac{B}{4}\phi^4 + \frac{B}{4}|\nabla\phi|^2 + \psi\ln c_1 \right).  \label{eq:F-renormalized}
\end{align}
Thus, the equilibrium free energy decouples into: $\mathcal{F}_{\text{eq}}[\phi,\psi]=\mathcal{F}_1[\phi]+\mathcal{F}_2[\psi]$.
Let us define the equilibrium free energy density $g_1(y)$ such that $\mathcal{F}_1=\int g_1(y)\,dV$.
Substituting the equilibrium solution for $\phi(y)$, given in Eqs.~(\ref{eq:sol_phi}) and (\ref{eq:sol_phi1}), into the free energy in Eq.~(\ref{eq:F-renormalized}),
we can then calculate the equilibrium free energy density $g_1(y)$.
Fig.~\ref{fig:surface-tension}(a) shows the typical plot of the free energy density $g_1(y)$ as a function of vertical distance $y$.
$g_1(y)$ shows a peak at the liquid-gas interface $y=h$.
The surface tension $\gamma$ is then defined to be the excess free energy~\cite{chaikin2000principles}, or the shaded area in Fig.~\ref{fig:surface-tension}(a),
which can be done by integrating $g_1(y)$, minus the bulk value, across the interface. In our case, we take the interface to be $y\in[h-10\xi,h+10\xi]$.

We can then plot the renormalized surface tension $\gamma$ as a function of the average concentration of colloidal particles $\left<\psi\right>=\frac{1}{\xi^2h}\int_0^h\psi(y)\,dy$ for different values of buoyant mass $G$ and coupling constant $C$, as shown in Fig.~\ref{fig:surface-tension}(b).
The plot shows that $\gamma$ decreases monotonically with increasing colloidal particle concentration $\left<\psi\right>$, 
and in the limit of zero particle ($\left<\psi\right>\rightarrow0$), the surface tension $\gamma$ approaches the bare value $\gamma_0=2B/3$.
This is consistent with the experimental observations reported  in~\cite{okubo1995surface}.
However, other experiments have also observed an initial decrease in the surface tension, followed by an increase in the surface tension as $\left<\phi\right>$ rises~\cite{dong2003surface,tanvir2012surface}.
This is because at higher concentrations, excluded volume interactions between the particles become important, which may cause correlation between the particles~\cite{zaccone2022colloid}.
In particular, the particles can form crystalline structure at the interface, which increase the capillary forces between the particles and results in higher surface tension~\cite{dong2003surface}.

From Fig.~\ref{fig:surface-tension}(b), we also observe that the surface tension is monotonically decreasing with decreasing buoyant mass $G$.
This can be explained by the fact that a lower buoyant mass (\emph{e.g.}, more negative) causes more particles to float towards the interface, 
resulting in a greater accumulation of particles near the interface.
Furthermore from Fig.~\ref{fig:surface-tension}(b), we also observe the surface tension to be monotonically decreasing with increasing coupling constant $C$.
This is because at higher $C$, the particles distort the interfacial profile $\phi(y)$ even more, leading to a greater impact on the surface tension.

Note that the surface tension is the equilibrium property of an interface, which is defined to be the interfacial energy per cross-sectional area.
Thus the surface tension remains the same in the purely diffusive regime (\emph{i.e.} $\mathbf{u}=0$), 

\subsection{Numerical simulations \label{sec:results}}

In 2D simulations, we consider a finite system size $L_x\times L_y$ with periodic boundary conditions at $x=0$ and $x=L_x$, 
as shown in Fig.~\ref{fig:2D}. 
We assume no-slip and no-flux boundary conditions at $y=0$ and $y=L_y$ to represent the walls at $y=0$ and $y=L_y$.
The liquid-gas interface is located at $h\simeq L_y/2$ and we assume $L_y$ to be large enough so that the analytic solution for the semi-infinite system in Section~\ref{sec:analytic} can be compared well to the numerical results. 

We initialize $\mathbf{u}$ to be zero and $\phi$ to be a step function, where $\phi=1$ for $y\le L_y/2$, and $-1$ for $y>L_y/2$.
We also initialize $\psi$ to be a step function, where $\psi=$ constant for $y\le L_y/2$, and $\psi=0$ for $y>L_y/2$.
The constant is chosen such that $\psi$ is normalized to $N_p$.
For hybrid particle-phase field simulations, we initialize the particles' positions to be randomly distributed across the region $y<L_y/2$.

The system reaches an equilibrium steady state after time $t>t_{\text{eq}}$ [see Eq.~(\ref{eq:t-eq})].
Fig.~\ref{fig:2D}(a,b) show the typical snapshots of the hybrid particle-phase field 2D simulations in steady state for $G>0$ (a) and $G<0$ (b).
As expected for $G>0$, the particles sediment towards the bottom, while for $G<0$, the particles float towards the liquid-air gas interface.
Here $C$ is chosen to be large enough so that $\sim99.98\,\%$ of the particles remain confined inside the liquid phase.
Some exceptions are, for example, the $2$ isolated particles in the top right corner in Fig.~\ref{fig:2D}(b), which are statistically negligible.
In hybrid particle-phase field simulations, a small, noisy fluid flow $\mathbf{u}$ still persists in the steady state. 
This is due to random fluctuations in the particles’ positions, which generate random stress in the Stokes equation.
However in continuum two-phase field model, this fluid flow $\mathbf{u}$ will decay to zero in the steady state.

We now compare the equilibrium interfacial profile $\phi(y)$ from numerical simulations with the perturbative solution from Section~\ref{sec:analytic}.
The results are shown in Fig.~\ref{fig:phi} for $G>0$ (a) and $G<0$ (b).
The simulations are here performed in 1D, as there is no distinction in the equilibrium profile between 1D and 2D simulations.
During the simulations, the height of the liquid-gas interface $h$ (defined as the value of $y$ where $\phi=0$) will move.
However, our perturbation theory cannot predict the final value of $h$, so we use it as a fitted parameter. 
We find the best fit for $h$ by minimizing the absolute average difference between the $\phi_1$ from perturbation theory
and the equivalent $\phi_1$ from the simulations.

The effect of strong coupling $C$ on $\phi$ is presented in Fig.~\ref{fig:phi}(a) for different values of $G>0$. 
The inset provides a zoomed-in view around $y=0$, where the concentration of the particles is the highest.
While the overall $\tanh$ shape is maintained, the value of $\phi$ is significantly altered near the highest concentration of particles at $y = 0$. 
Fig.~\ref{fig:phi}(b) shows similar plots for $G\le0$.
The inset shows the zoomed-in view around the interface at $y=h$, where the concentration of particles is the highest.
For $G=0$, we observe that there is very little deviation from the classical $\tanh$ profile while for $G<0$ and large $C$, we again observe a peak near the interface.
In both figures, the magnitude of the peak increases with increasing $C$. 

Fig~\ref{fig:phi} shows a very clear impact on the interfacial profile $\phi(y)$ from strong coupling with particles' positions $\{\mathbf{r}_i\}$ 
(or equivalently particle distribution $\psi(y)$). 
This is in direct contrast to~\cite{van2006diffuse,zong2020modeling,ma2012modeling,zhao2017modeling}, 
where it was argued that the interfacial profile $\phi(y)$ is independent of the presence of particles. 
However, these previous models focused on adsorption of surfactant particles at the interface, 
which can be characterized as operating within a weak-coupling regime.

Fig.~\ref{fig:psi} shows the particle distribution $\psi(y)$ at equilibrium for two-phase field simulation, hybrid particle-phase field simulation, and analytic solution from perturbation theory. 
It is clear to see that $G$ is responsible for the shape of the $\psi$ distributions whereas $C$ is responsible for confinement of particles into the region $y<h$.
When $G = 0$, the particles do not shift from a uniform distribution, as shown in Fig.~\ref{fig:psi} bottom left. 
This is close to how particles will behave under large sedimentation lengths $k_{\text{B}}T/(\tilde{m}g)$. 

\subsection{Discussion \label{sec:results}}

Fig.~\ref{fig:psi} demonstrates an excellent agreement between numerical simulations and perturbation theory for the particle distribution $\psi(y)$.
Likewise, Fig.~\ref{fig:phi} shows a reasonably good agreement between the simulated interfacial profile $\phi(y)$ and the analytic solution $\phi(y)$, 
with percentage error of around $~0.7\%$.

This error arises from the finite size effect of the simulation.
While perturbation theory is performed on a semi-infinite system $y\in[0,\infty)$,  
the simulations are performed in a finite system $y\in[0,L_y]$.
As a result, during the simulation the height of the interface $h$ might shift from its initial value $h=L_y/2$.
The shift in $h$ causes the bulk value $\phi(y)$ in the gaseous phase ($y>h$) to move away from the binodal value ($\phi=-1$).
Consequently, this results in an apparent vertical shift in the numerical $\phi(y)$, compared to the analytic $\phi(y)$, shown in Fig.~\ref{fig:phi}.
Additionally, the error may also originate from the lattice discretization $\Delta y$, which has a greater impact when $G<0$.


As demonstrated in Section IIIB, the presence of particles near the interface can distort the interfacial profile $\phi(y)$, which subsequently decreases the surface tension. This reduction of surface tension is also observed in surfactant-laden interface~\cite{xu2023improved}, however, we should emphasize that the underlying mechanism is different.
In the case of surfactant, the surfactant molecules are adsorped into the interface. 
This adsorption gives rise to a coupling term inside the free energy functional, which is proportional to $\psi|\nabla\phi|^2$.
This is different from our coupling term, which is proportional to $\psi\phi$.

\section{Conclusion}

In conclusion, we have developed two complementary models to simulate the dynamics of dilute colloidal sedimentation and flotation near a liquid-gas interface. 
These models assume purely local interactions between the colloidal particles and the fluid, simplifying previous approaches~\cite{verberg2005modeling}.
Despite this simplification, the fundamental physics of the system remains accurately represented.

The assumption of local interaction allows for an analytical solution of the equilibrium interfacial profile, which shows excellent agreement with numerical simulations. 
This research underscores the significant impact of particle dynamics on the liquid-gas interface, including changes in interfacial tension, which we have successfully derived mathematically using a perturbative approach. 
Specifically, our model demonstrates that surface tension decreases monotonically with increasing particle concentration and decreasing buoyant mass of the particles, in agreement with many experimental observations~\cite{okubo1995surface,dekker2023difference}. 
However, at higher concentrations, experimental observations sometimes reveal more complex, non-monotonic behaviour of surface tension due to finite size effects of the particles~\cite{dong2003surface}, a factor not considered in our model.

In the dilute limit,  our model provides a robust framework for capturing the complex dynamics of colloidal particles in fluid systems, which can be potentially extended to suspensions of active colloids near a fluid-fluid interface~\cite{bishop2017active,stebe2022active,zottl2016review}.

\begin{acknowledgments}
AJH acknowledges EPSRC DTP studentship no. 2739112.
ET acknowledges funding from EPSRC grant no. EP/W027194/1. 
\end{acknowledgments}

\appendix

\section{Derivation of the free energy from Flory-Huggins theory \label{app:lattice}}

In this Appendix, we will derive the free energy functional in Eq.~(\ref{eq:free-energy}) of the main text through explicit coarse-graining of the underlying sub-lattice system, also known as the Flory-Huggins theory. 
Alternatively, the free energy can also be derived from the partition function using density functional theory~\cite{hughes2014introduction}, 
another approach which we are not going to use here.
The derivation below is performed in 2D, although extension to 3D should be straightforward.

\begin{figure}
    \centering
    \includegraphics[width=0.9\linewidth]{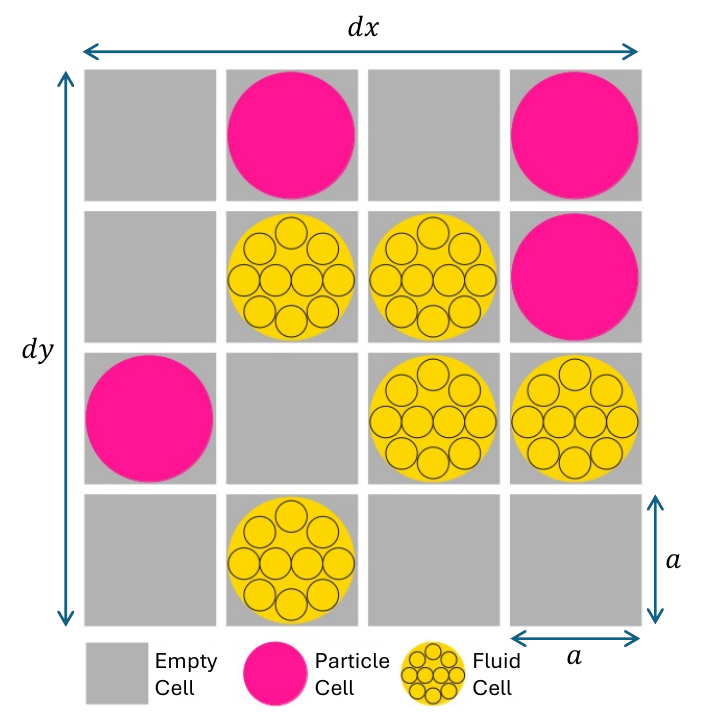}
    \caption{An area element $dx\times dy$ is divided into a sub-lattice of $N$ cells of size $a\times a$, where $a$ is the diameter of the particles.
    Each cell can be: empty, occupied by a single particle, or occupied by the fluid.}
    \label{fig:lattice}
\end{figure}

Our system is defined as a 2D box, sized $L_x\times L_y$ containing colloidal particles and fluid molecules. 
The box is further divided into area elements sized $dx\times dy$. 
These area elements are then divided further into a sub-lattice of $N$ cells of size $a\times a$, 
where $a$ is the diameter of the particles (see Figure \ref{fig:lattice}).
Each cell can be: empty, occupied by a single particle, or occupied by the fluid molecules. 
Since fluid molecules and colloidal particles have very different diameters, we assume that when the cell is to be occupied by the fluid, 
it is filled with as many molecules as needed to make the total area equal to that of the particle, and that this area acts as one fluid `particle'. 
This assumption is used in \cite{chalmers2017dynamical} and proved in \cite{vancea2008front}. 

Let us denote $N_e$ to be the number of empty cells, $N_p$ to be the number particle-filled cells, and $N_f$ to be the number of fluid-filled cells,
obeying $N_e+N_p+N_f=N$.
The free energy of this area element (which is a lattice of $N$ cells) can be written as: $F = U - TS$, where $U$ is the potential energy, $T$ is temperature and $S$ is the entropy.

The potential energy has two contributions: interactions between the constituents and interactions with the external fields, of which we only consider gravity. 
The former comes from neighbouring pairs of fluid-fluid and fluid-particle cells. 
A fluid-fluid pair interacts with energy $-\epsilon_{ff}<0$ and a fluid-particle pair interacts with energy $-\epsilon_{fp}<0$.
The interaction energies are both negative to indicate attraction between fluid-fluid pair and between fluid-particle pair
(we assume that there is no interaction between two particles).
Let us denote $z$ to be the number of nearest neighbour cells (\emph{i.e.} $z=4$ in 2D and $z=6$ in 3D). 
The probability of having a fluid neighbour is $N_f/N$ and the probability of having a particle neighbour is $N_p/N$.
Therefore the potential energy is:
\begin{equation}
U = -\frac{1}{2}\epsilon_{ff}zN_f \frac{N_f}{N}- \epsilon_{fp}zN_f\frac{N_p}{N} + \tilde{m}gyN_p. \label{eq:U}
\end{equation}
The factor of half in the first term comes from double counting of the fluid-fluid pairs.
The last term in (\ref{eq:U}) is the gravitational potential energy of the particles.

The entropy is calculated from the formula $S = k_{\text{B}}\ln\Omega$, where $\Omega$ is the number of microstates, which is equal to the number of permutations we can put particles and fluid into $N$ cells:
\begin{align}
    \Omega = \frac{N!}{N_e!N_p!N_f!}.
\end{align}
With use of Stirling's approximation $\ln N! = N\ln N - N$ and $N_e=N-N_p-N_f$, we can then write the free energy per unit cell:
\begin{align}
    \frac{F}{N} &= -\frac{1}{2}\epsilon_{ff}z\frac{N_f}{N} \frac{N_f}{N}- \epsilon_{fp}z\frac{N_f}{N}\frac{N_p}{N} + \tilde{m}gy\frac{N_p}{N} \nonumber\\
    		      &+ k_{\text{B}}T\left[ \frac{N_p}{N}\ln\left(\frac{N_p}{N}\right) + \frac{N_f}{N}\ln\left(\frac{N_f}{N}\right) \right. \nonumber\\ 
		      & \left.+ \left(1 - \frac{N_p}{N}-\frac{N_f}{N}\right)\ln\left(1 - \frac{N_p}{N}-\frac{N_f}{N}\right) \right].
    \label{eq:free-energy-1}
\end{align}
The next step is to expand the free energy around the critical point.
In a pure gaseous phase, we have $N_f=0$ and $N_p=0$ since all cells are empty.
In a pure liquid phase, we have $N_f=N-N_p$, \emph{i.e.} all cells are occupied by either fluid or particles.
At the critical point we hypothesize that $N_f=(N-N_p)/2$ (this choice will eliminate the cubic $\sim\phi^3$ term from the free energy). 
We can then expand $N_f+N_p/2$ around the critical point by writing:
\begin{equation}
\frac{N_f}{N}+\frac{N_p}{2N} = \frac{1}{2} + \phi, \label{eq:phi}
\end{equation}
where $\phi$ is a small expansion parameter.
From the definition in (\ref{eq:phi}), we can also interpret $\phi$ as the combined number density of fluid and particles (rescaled and weighted by some factors).
From Eq.~(\ref{eq:phi}), we can also infer that $\phi<0$ corresponds to the gaseous phase while $\phi>0$ corresponds to the liquid phase.
We also define the particle number density to be:
\begin{equation}
\psi = \frac{N_p}{Na^2}. \label{eq:psi}
\end{equation}
Note that $\psi$ has the dimension of one over area whereas $\phi$ is dimensionless.
We can then substitute Eqs.~(\ref{eq:phi}-\ref{eq:psi}) into Eq.~(\ref{eq:free-energy-1}) and Taylor expand for small $\phi$ and $\psi$ to get:
\begin{align}
\frac{F}{N} &= \tilde{m}gya^2\psi - \frac{1}{2}\epsilon_{ff}z\left(\frac{1}{4}a^4\psi^2-a^2\psi\phi+\phi^2\right) \nonumber\\
&- \epsilon_{fp}z\left(-\frac{1}{4}a^4\psi^2 + a^2\psi\phi \right) \nonumber\\
&+ k_{\text{B}}T \left[ a^2\psi\ln(a^2\psi) + (2+2a^2\psi+2a^4\psi^2)\phi^2 + \frac{4}{3}\phi^4 \right. \nonumber\\
& \left. +\frac{1}{2}a^4 \psi^2 + \frac{1}{6}a^6\psi^3 + \frac{1}{12}a^8\psi^4 \right], \label{eq:free-energy-2}
\end{align}
where we have ignored the linear and constant terms in $F$, since they do not affect the $\phi$- and $\psi$-dynamics in Eqs.~(\ref{eq:phidot}-\ref{eq:psidot}).
To get the free energy density, we divide (\ref{eq:free-energy-2}) by $a^2$ and make further assumption that the particle density is much smaller than the combined density $a^2\psi\ll\phi$ (dilute limit).
Loosening the assumption of $a^2\psi\ll\phi$ to allow up to second order of $\psi$ will result in more coupling terms, like those seen in the surfactant free energy~\cite{liu2010phase}.
The result is:
\begin{align}
\frac{F}{Na^2} &= \underbrace{\left(\frac{2k_{\text{B}}T}{a^2}-\frac{\epsilon_{ff}z}{2a^2}\right)}_{\alpha/2}\phi^2 + \underbrace{\frac{4k_{\text{B}}T}{3a^2}}_{\beta/4}\phi^4 
+ \tilde{m}gy\psi \nonumber\\
&+ k_{\text{B}}T \psi\ln(a^2\psi) - \underbrace{\left(\epsilon_{fp}z - \frac{1}{2}\epsilon_{ff}z\right)}_{c}\psi\phi. \label{eq:free-energy-flory}
\end{align}
The free energy density above has exactly the same form as in Eq.~(\ref{eq:free-energy}) of the main text, except that we did not get the gradient term $\propto |\nabla\phi|^2$ from Flory-Huggins theory.
This is because the free energy in Eq.~(\ref{eq:free-energy-flory}) is purely local to a particular area element $dx\times dy$.
To obtain the gradient terms such as $|\nabla\phi|^2$, we need to consider the interactions between one area element $dx\times dy$ with the neighbouring area elements.
Alternatively, the squared gradient term can also be added phenomenologically as we have done in this paper.
Note, there could be more than one possible gradient term as shown in~\cite{zhong2015sessile,cogswell2011thermodynamic}.

We also note that in this derivation, we have neglected the interactions between colloidal particles.
However, the free energy can, in principle, be generalized to account for these interactions. 
If attractive forces are present, it is likely that the particles will phase separate, forming dense and dilute regions within the liquid phase of the fluid.

\begin{figure}
    \centering
    \includegraphics[width=1\linewidth]{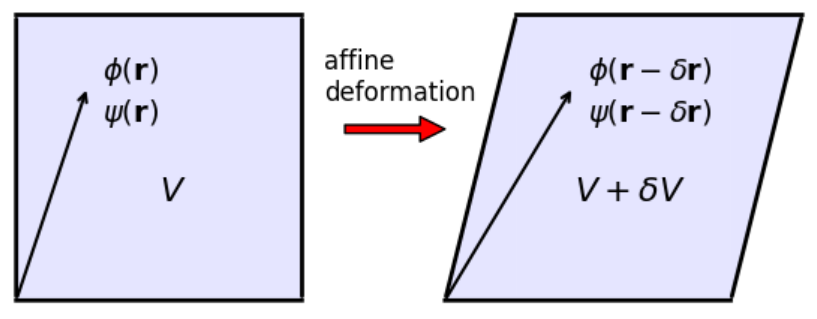}
    \caption{To derive the elastic stress tensor $\underline{\underline{\boldsymbol{\sigma}}}$, we perform an affine deformation $\mathbf{r}\rightarrow\mathbf{r}+\delta\mathbf{r}$ on the system. The change in the free energy is then given by the integral of the elastic stress tensor times the strain tensor: $\delta\mathcal{F}=\int \sigma_{\alpha\beta}\partial_\beta\delta r_\alpha\,dV$.}
    \label{fig:affine}
\end{figure}

\section{Derivation of the stress tensor \label{app:stress}}

In this Appendix, we derive the formula for the elastic stress tensor $\underline{\underline{\boldsymbol{\sigma}}}$ from the free energy functional of the form~\cite{markovich2019chiral}:
\begin{equation}
\mathcal{F}[\boldsymbol{\Phi}] = \int g(\boldsymbol{\Phi},\partial_\alpha\boldsymbol{\Phi})\,dV,
\end{equation}
where $\boldsymbol{\Phi}=(\phi,\psi,\dots)$.
$g$ is the free energy density which depends on $\boldsymbol{\Phi}$ and its derivatives $\partial_\alpha\boldsymbol{\Phi}$.
$\partial_\alpha$ indicates partial derivative with respect to spatial coordinate $\alpha$, where $\alpha=x,y,$ or $z$.
In our model, $g$ does not depend explicitly on $\partial_\alpha\psi$.
Suppose that our fluid $\boldsymbol{\Phi}(\mathbf{r})$ is confined inside a rectangular box of volume $V$, as shown in Fig.~\ref{fig:affine}.
Now we can deform the fluid affinely (\emph{e.g.} by shearing the box) through some infinitesimal strain $\delta\mathbf{r}$.
In other words, we displace every small patch of fluid material from $\mathbf{r}$ to $\mathbf{r}+\delta\mathbf{r}$.
Under this affine deformation, $\boldsymbol{\Phi}(\mathbf{r})$ and $V$ transform as:
\begin{align}
\boldsymbol{\Phi} &\rightarrow \boldsymbol{\Phi}(\mathbf{r}-\delta\mathbf{r}) = \boldsymbol{\Phi}(\mathbf{r}) - \delta\mathbf{r}\cdot\nabla\boldsymbol{\Phi} \\
V &\rightarrow V+\delta V.
\end{align}
We may define $\delta\boldsymbol{\Phi}=-\delta\mathbf{r}\cdot\nabla\boldsymbol{\Phi}$.
Now the total change in the free energy due to this affine deformation is given by:
\begin{align}
\delta \mathcal{F} =& \mathcal{F}[\boldsymbol{\Phi}+\delta\boldsymbol{\Phi}] - \mathcal{F}[\boldsymbol{\Phi}] \\
 =& 
\int_{V} \bigg\{ \delta\boldsymbol{\Phi}\cdot\frac{\partial g}{\partial\boldsymbol{\Phi}} + (\partial_\alpha\delta\boldsymbol{\Phi})\cdot\frac{\partial g} {\partial(\partial_\alpha\boldsymbol{\Phi})} \bigg\}\,dV \nonumber\\
&+ \int_{\delta V} g(\boldsymbol{\Phi},\partial_\alpha\boldsymbol{\Phi}) \, dV \label{eq:dF1}
\end{align}
Note that we can write the last term as:
\begin{equation}
\int_{\delta V} g(\boldsymbol{\Phi},\partial_\alpha\boldsymbol{\Phi}) \, dV  = \oint_{\partial V} g(\boldsymbol{\Phi},\partial_\alpha\boldsymbol{\Phi}) \, \delta\mathbf{r}\cdot d\mathbf{S},
\end{equation}
since when we displace a surface element $d\mathbf{S}$ by $\delta\mathbf{r}$, the volume covered by this travelling surface element is $\delta\mathbf{r}\cdot d\mathbf{S}$.
Using integration by parts, incompressibility condition $\partial_\alpha\delta r_\alpha=0$ and $\delta\boldsymbol{\Phi}=-\delta\mathbf{r}\cdot\nabla\boldsymbol{\Phi}$ on (\ref{eq:dF1}), we get:
\begin{align}
\delta \mathcal{F}
=& \oint_{\partial V} \left\{ \left(g - \boldsymbol{\Phi}\cdot\frac{\delta\mathcal{F}}{\delta\boldsymbol{\Phi}}\right)\delta_{\alpha\beta} - 
(\partial_\alpha\boldsymbol{\Phi})\cdot\frac{\partial g}{\partial(\partial_\beta\boldsymbol{\Phi})}  \right\} \delta r_\alpha  \, dS_\beta  \nonumber\\
&+ \int_{V}  \left\{\boldsymbol{\Phi}\cdot \partial_\alpha\left(\frac{\delta\mathcal{F}}{\delta\boldsymbol{\Phi}}\right)\right\} \delta r_\alpha \, dV. \label{eq:dF2}
\end{align}
From the first law of thermodynamics, the change in the free energy is also equal to:
$\delta \mathcal{F} = \delta W - \delta Q - T\delta S$,
where $\delta W$ is the work done on the system, $\delta Q$ is the heat dissipated into the environment and $\delta S$ is the increase in the system's entropy.
The heat dissipated into the environment causes the entropy of the environment (or heat reservoir) $S_r$ to increase.
Assuming the reservoir to be frictionless, we can write $\delta Q = T\delta S_r$.
Thus, the change in the free energy is:
$\delta\mathcal{F} = \delta W - T (\delta S + \delta S_r)$.
Note that $\delta S + \delta S_r$ is the change in the total entropy.
For all affine deformations, the process is time-reversible and should not contribute to the total entropy.
Therefore, the change in the free energy is simply the work done on the system by the external forces: $\delta\mathcal{F} = \delta W$.

There are two types of external forces: 1) external surface force $\mathbf{F}^{\text{surface}}$ which is acting on the walls of the container, and
2) external body force $\mathbf{F}^{\text{body}}$ which is acting on the bulk of the fluid.
The force acting on the walls by the external force is $\mathbf{F}^{\text{surface}}$ and
the force acting on the walls by the fluid is $-\oint_{\partial V}\sigma_{\alpha\beta}\,dS_\beta$.
Since we assume mechanical equilibrium throughout, the net force on the walls has to be zero:
\begin{equation}
F^{\text{surface}}_\alpha - \oint_{\partial V}\sigma_{\alpha\beta}\,dS_\beta = 0, \quad\text{for } \alpha=x,y,z.
\end{equation}
Similarly, the external force acting on the bulk of the fluid $\mathbf{F}^{\text{body}}$
is balanced by the force from the fluid itself $\int\partial_\beta\sigma_{\alpha\beta}\,dV$:
\begin{equation}
F^{\text{body}}_\alpha + \int\partial_\beta\sigma_{\alpha\beta}\,dV = 0, \quad\text{for } \alpha=x,y,z.
\end{equation}
Thus the work done by the external forces, $\delta W = \mathbf{F}\cdot\delta\mathbf{r}$, can be written as:
\begin{align}
\delta W 
&= \mathbf{F}^{\text{surface}}\cdot \delta\mathbf{r} + \mathbf{F}^{\text{body}}\cdot\delta\mathbf{r} \\
&= 
\oint_{\partial V} \sigma_{\alpha\beta} \delta r_\alpha \, dS_\beta  - 
\int_{V} (\partial_\beta\sigma_{\alpha\beta}) \delta r_\alpha \, dV \label{eq:dW}\\
&= \int_V \sigma_{\alpha\beta} \partial_\beta \delta r_\alpha \, dV
\end{align}
which is the integral of the stress times the strain tensor $\partial_\beta\delta r_\alpha$.
Now we can then equate (\ref{eq:dF2}) to (\ref{eq:dW}) since $\delta\mathcal{F} = \delta W$.
Comparing the surface term, we get the elastic stress tensor:
\begin{equation}
\sigma_{\alpha\beta} = \left(g - \boldsymbol{\Phi}\cdot\frac{\delta\mathcal{F}}{\delta\boldsymbol{\Phi}} \right)\delta_{\alpha\beta} 
 -  (\partial_\alpha\boldsymbol{\Phi})\cdot\frac{\partial g}{\partial(\partial_\beta\boldsymbol{\Phi})},  \label{eq:sigma-general}
\end{equation}
where we have also identified the isotropic pressure to be $p=\boldsymbol{\Phi}\cdot\frac{\delta\mathcal{F}}{\delta\boldsymbol{\Phi}}-g$.
In our model, $g$ does not depend explicitly on $\partial_\alpha\psi$ and thus $\psi$ only contributes to the isotropic pressure explicitly 
(although $\psi$ can modify $\phi$ and contribute to the anisotropic stress $\sigma_{\alpha,\beta\neq\alpha}$ indirectly through $\phi$).
Equating the volume term in (\ref{eq:dF2}) and (\ref{eq:dW}), we get the force density:
\begin{equation}
\mathbf{f} = -\boldsymbol{\Phi}\cdot\nabla\frac{\delta\mathcal{F}}{\delta\boldsymbol{\Phi}}
= -\phi\nabla\frac{\delta\mathcal{F}}{\delta\phi} - \psi\nabla\frac{\delta\mathcal{F}}{\delta\psi} \label{eq:f-general}
\end{equation}
One can also verify that $f_\alpha=\partial_\beta\sigma_{\alpha\beta}$ in (\ref{eq:sigma-general}-\ref{eq:f-general}).

\section{Dimensionless equations \label{app:dimensionless}}

In this Appendix, we will recast the equations of motion from Section~\ref{sec:two-phase} and \ref{sec:hybrid} in their dimensionless form. 
First, without loss of generality, we can set $\alpha=-\beta$.
We then define the interfacial width $\xi=\sqrt{2\kappa/\beta}$ to be the unit of length, $\tau=\lambda\xi^2/(k_{\text{B}}T)$ to be the unit of time, and $k_{\text{B}}T$ to be the unit of energy.
Roughly speaking $\tau$ is the timescale for the particles to diffuse a distance $\xi$.
We define the dimensionless position $\bar{\mathbf{r}}$, dimensionless time $\bar{t}$, dimensionless particle density $\bar{\psi}$, and dimensionless fluid velocity $\bar{\mathbf{u}}$ to be:
\begin{equation}
\mathbf{r} = \xi\bar{\mathbf{r}}, \quad t = \tau\bar{t}, \quad \psi = \frac{1}{\xi^3}\bar{\psi}, \quad\text{and}\quad \mathbf{u} = \frac{\xi}{\tau}\bar{\mathbf{u}}.
\label{eq:dimensionless-variables}
\end{equation}
We also define the dimensionless pressure $\bar{p}$, dimensionless force density $\bar{\mathbf{f}}$ and renormalized viscosity $\bar{\eta}$ to be:
\begin{equation}
\bar{p} = \frac{\xi^3}{k_{\text{B}}T}p, \quad \bar{\mathbf{f}} = \frac{\xi^4}{k_{\text{B}}T}\mathbf{f}, \quad\text{and}\quad \bar{\eta} = \frac{\xi}{\lambda}\eta. \label{eq:dimensionless-variables-2}
\end{equation}
Now substituting Eq.~(\ref{eq:dimensionless-variables}-\ref{eq:dimensionless-variables-2}) into (\ref{eq:phidot}-\ref{eq:Stokes}), we get the dimensionless form of the two-phase field model:
\begin{align}
\frac{\partial\phi}{\partial\bar{t}} + (\bar{\mathbf{u}}\cdot\bar{\nabla})\phi &= \bar{M}_\phi\bar{\nabla}^2\left(\frac{\delta\bar{\mathcal{F}}}{\delta\phi}\right) 
\label{eq:phidot-dimensionless} \\
\frac{\partial\bar{\psi}}{\partial\bar{t}} + (\bar{\mathbf{u}}\cdot\bar{\nabla})\bar{\psi} &= 
\bar{\nabla}\cdot\left(\bar\psi\nabla\frac{\delta\bar{\mathcal{F}}}{\delta\bar\psi} \right) \label{eq:psidot-dimensionless} \\
0 &= -\bar{\nabla}\bar{p} + \bar{\eta}\bar{\nabla}^2\bar{\mathbf{u}} + \bar{\mathbf{f}}, \label{eq:Stokes-dimensionless}
\end{align}
where $\bar{\mathcal{F}}=\mathcal{F}/k_{\text{B}}T$ is the dimensionless free energy:
\begin{align}
\bar{\mathcal{F}}[\phi,\bar{\psi}] &= \int d\bar{V} \bigg( -\frac{\bar{B}}{2}\phi^2 + \frac{\bar{B}}{4}\phi^4 + \frac{\bar{B}}{4}|\bar{\nabla}\phi|^2 \nonumber\\
&+ \bar{G}\bar{y}\bar{\psi} + \bar{\psi}\ln\bar{\psi} - \bar{C}\bar{\psi}\phi \bigg),
\end{align}
and we have defined the dimensionless quantities $\bar{M}_\phi,\bar{B},\bar{C},$ and $\bar{G}$ to be:
\begin{equation}
\bar{M}_\phi = \frac{M_\phi\lambda}{\xi^3}, \,\, \bar{B} = \frac{\beta\xi^3}{k_{\text{B}}T}, \,\, \bar{C} = \frac{c}{k_{\text{B}}T}, \,\, \text{and} \,\,
\bar{G} = \frac{\tilde{m}g\xi}{k_{\text{B}}T}.
\end{equation}
Similarly, we can also write the Langevin equation (\ref{eq:rdot}) in its dimensionless form:
\begin{equation}
\frac{d\bar{\mathbf{r}}_i}{d\bar{t}} = \bar{\mathbf{u}}(\bar{\mathbf{r}}_i,\bar{t}) - \bar{G} \hat{\mathbf{y}} + \bar{C}\bar{\nabla}\phi(\bar{\mathbf{r}}_i,\bar{t})
+ \sqrt{2} \bar{\boldsymbol{\zeta}}_i(\bar{t}), \label{eq:rdot-dimensionless}
\end{equation}
where $\bar{\boldsymbol{\zeta}}_i=\sqrt{\tau}\boldsymbol{\zeta}_i$ is the dimensionless Gaussian white noise with zero mean and unit variance.
All numerical and analytical results in the main text are presented in the dimensionless form, with all the bars removed for ease of notation. 

\section{Numerical methods \label{app:numerics}}

To solve the $\phi$- and $\psi$-dynamics in Eqs.~(\ref{eq:phidot}-\ref{eq:psidot}), we discretize the time into timesteps $\Delta t$ and space into lattice grid $\Delta x\times\Delta y$, so that $L_x=N_x\Delta x$ and $L_y=N_y\Delta y$, where $N_x$ and $N_y$ are the number of lattice points in the $x$- and $y$-direction.
Ideally $\Delta t$, $\Delta x$ and $\Delta y$  have to be small, and in simulations, we use $\Delta t=0.001$ and $\Delta x=\Delta y=0.5$, unless stated otherwise.
Eqs.~(\ref{eq:phidot}-\ref{eq:psidot}) are then solved using standard central finite difference for the spatial derivatives and Euler update for the time derivative.
By first writing Eqs.~(\ref{eq:phidot}-\ref{eq:psidot})  as continuity equations $\partial\phi/\partial t+\nabla\cdot\mathbf{J}_\phi=0$ and $\partial\psi/\partial t+\nabla\cdot\mathbf{J}_\psi=0$, the no-flux conditions at $y=0$ and $y=L_y$ are imposed by fixing:
\begin{align}
\left.\mathbf{J}_\phi\cdot\hat{\mathbf{y}}\right|_{y=0} &= \left.\mathbf{J}_\phi\cdot\hat{\mathbf{y}}\right|_{y=L_y} = 0 \\
\left.\mathbf{J}_\psi\cdot\hat{\mathbf{y}}\right|_{y=0} &= \left.\mathbf{J}_\psi\cdot\hat{\mathbf{y}}\right|_{y=L_y} = 0,
\end{align}
where $\hat{\mathbf{y}}$ is a unit vector in the positive $y$-direction.
We also impose a `neutral wetting' conditions on $\phi$ at the walls by fixing~\cite{kruger2017LB}:
\begin{equation}
\left.\frac{\partial\phi}{\partial y}\right|_{y=0} = \left.\frac{\partial\phi}{\partial y}\right|_{y=L_y} = 0. \label{eq:neutral-wetting}
\end{equation}
To solve the particles' dynamics in Eq.~(\ref{eq:rdot}), we again discretize the time into the same timesteps $\Delta t$,
but the particles' positions $\{\mathbf{r}_i\}$ now live off-lattice. 
Eq.~(\ref{eq:rdot}) is solved using standard Euler-Maruyama scheme with periodic boundary conditions at $x=0$ and  $x=L_x$, and no-flux boundary conditions at $y=0$ and $y=L_y$.

The Stokes equation in (\ref{eq:Stokes}) is solved using spectral method as follow.
First, $\mathbf{u}$ is periodic along $x$ such that $\mathbf{u}(x,y)=\mathbf{u}(x+L_x,y)$ for all $x$ and $y$.
From no-slip boundary conditions, we must have $\mathbf{u}(x,y=0)=\mathbf{u}(x,y=L_y)=0$ for all $x$.
Thus, we can expand $\mathbf{u}$ in terms of the complex exponentials along $x$ and in terms of sines along $y$:
\begin{equation}
\mathbf{u}(x,y) = \sum_{n,m} \tilde{\mathbf{u}}_{n,m} e^{i\frac{n2\pi x}{L_x}}\sin\left(\frac{m\pi y}{L_y}\right),  \label{eq:mixed-transform}
\end{equation}
where $n$ and $m$ are integers.
$\{\tilde{\mathbf{u}}_{n,m}\}$ are a set of Fourier coefficients, which can be found using orthogonality and completeness relation of the sines and complex exponentials.
In Eq.~(\ref{eq:mixed-transform}), we have a mix of sine and Fourier transform, which is not currently implemented in many numerical libraries (such as NumPy).
To fix this issue, we shall extend the domain for $\mathbf{u}(x,y)$ from $y\in[0,L_y]$ to $y\in[0,2L_y]$ (the domain for $x$ remains the same). 
We can then write the summation in (\ref{eq:mixed-transform}) as:
\begin{equation}
\mathbf{u}(x,y) = \sum_{\mathbf{k}} \tilde{\mathbf{u}}_{\mathbf{k}} e^{ik_x x}e^{ik_y y},  \label{eq:Fourier-transform}
\end{equation}
with the condition that $\mathbf{u}$ is odd with respect to $y\rightarrow -y$ and $\mathbf{u}$ is also periodic with period $2L_y$ along $y$:
\begin{equation} 
\mathbf{u}(x,y) = \mathbf{u}(x,y+2L_y) \,\,\text{and}\,\,  \mathbf{u}(x,-y)=-\mathbf{u}(x,y), \label{eq:symmetry}
\end{equation}
for all $x$ and $y$.
The wavector $\mathbf{k}=(k_x,k_y)$ is defined to be: 
\begin{equation}
k_x = \frac{2\pi}{L_x}n, \quad\text{and}\quad k_y = \frac{\pi}{L_y} m,
\end{equation}
where $n$ and $m$ are integers.
Applying the Fourier transform in (\ref{eq:Fourier-transform}), the incompressible Stokes equation (\ref{eq:Stokes-dimensionless}) becomes:
\begin{equation}
\tilde{u}_{\alpha,\mathbf{k}} = 
\begin{cases}
\frac{1}{\eta}\left(\frac{\delta_{\alpha\beta}}{k^2} - \frac{k_\alpha k_\beta}{k^4} \right) \tilde{f}_{\beta,\mathbf{k}}, &\quad\text{for }k\neq0 \\
0, &\quad\text{for }k=0
\end{cases},
\label{eq:utilde}
\end{equation}
where $\tilde{\mathbf{f}}_\mathbf{k}$ is the Fourier transform of the force density $\mathbf{f}(\mathbf{r})$, as given in Eq.~(\ref{eq:force-density}).
From Eq.~(\ref{eq:force-density}), we can calculate $\mathbf{f}(x,y)$ for $y\in[0,L_y]$. 
To find $\mathbf{f}(x,y)$ in the domain $y\in[L_y,2L_y]$, we exploit the symmetry given in (\ref{eq:symmetry}):
\begin{equation}
\mathbf{f}(x,2L_y-y) = -\mathbf{f}(x,y),\quad\text{for } y\in[0,L_y].
\end{equation}
We then use this force to calculate $\mathbf{u}(x,y)$ in the domain $y\in[0,2L_y]$ and we disregard $y\in[L_y,2L_y]$ when plotting the results.
The code for this simulation is available in~\cite{code}. 

\section{Estimation of physical parameters \label{app:parameters}}

In the simulations, we fix the unit of length to be $\xi=10\,\text{nm}$ so that the height of the interface is around $640\,\text{nm}$.
The diffusion constant of a nanoparticle is approximately $D\simeq1\,\mu\text{m}^2/\text{s}$.
Thus the unit of time for our simulation is $\tau=\xi^2/D\simeq0.1\,\text{ms}$.
The equilibration time is $t_{eq}=h\lambda/(\tilde{m}g)=\bar{h}\tau/\bar{G}$ [\emph{cf.} Eq.~(\ref{eq:t-eq})].
For $\bar{G}=0.05$, this gives equilibration time of around $1\,\text{s}$.
In most of our simulations, we fix the average concentration to be $\left<\bar{\psi}\right>=4$, which corresponds to $0.004/\text{nm}^3$ in physical units.
The bare surface tension is $\gamma_0=2Bk_BT/(3\xi^2)\simeq0.072 \,\text{Nm}^{-1}$ in physical units.
The rest of the parameters are summarized in Table~\ref{tab:params}.

\begin{table}[h]
\begin{tabular}{c|c}
Parameter& Value/Range\\
\hline
$\xi$ & $0.1\,\text{mm}$ \\
$\tau$ & $1\,\text{hr}$ \\
$\left<\bar{\psi}\right>$ & 2 (dimensionless) \\
$\bar G$ & -0.05, -0.01, 0.0, 0.01, 0.05, 0.1(dimensionless) \\
$\bar C$ & 3, 5 (dimensionless) \\
$\bar B$ & 500 (dimensionless) \\
$\bar{M}_{\phi}$ & 0.002 (dimensionless) \\
$\bar{\eta}$ & 10 (dimensionless) \\
\end{tabular}
\caption{\label{tab:params}
A table detailing the values of all parameters used.}
\end{table}

\bibliography{manuscript}

\end{document}